# A METHOD TO IDENTIFY POTENTIAL AMBIGUOUS MALAY WORDS THROUGH AMBIGUITY ATTRIBUTES MAPPING: AN EXPLORATORY STUDY

Hazlina Haron[1,2] and Abdul Azim Abd. Ghani [3]

[1] Faculty of Computer Science and Information Technology
Universiti Putra Malaysia
Serdang, Selangor Malaysia
delinn1612@yahoo.com

[2] School of Computing, College of Arts and Sciences
Universiti Utara Malaysia
Sintok, 06010, Kedah, Malaysia
zlina1108@uum.edu.my

[3]Faculty of Computer Science and Information Technology
Universiti Putra Malaysia
Serdang, Selangor Malaysia
azim@fsktm.upm.edu.my

## ABSTRACT

*We describe here a methodology to identify a list of ambiguous Malay words that are commonly being used in Malay documentations such as Requirement Specification. We compiled several relevant and appropriate requirement quality attributes and sentence rules from previous literatures and adopt it to come out with a set of ambiguity attributes that most suit Malay words. The extracted Malay ambiguous words (potential) are then being mapped onto the constructed ambiguity attributes to confirm their vagueness. The list is then verified by Malay linguist experts. This paper aims to identify a list of potential ambiguous words in Malay as an attempt to assist writers to avoid using the vague words while documenting Malay Requirement Specification as well as to any other related Malay documentation. The result of this study is a list of 120 potential ambiguous Malay words that could act as guidelines in writing Malay sentences.*

## 1. INTRODUCTION

Requirement Specification is a document that acts as a medium between system developer and users. Users specified their systems' functional needs in a technical documentation. The specification would then be referred by system analysts in the process of developing the requested system. Requirement Specification usually uses natural language, due to its' flexibility and easy to understand. However, natural language has its own disadvantages such as, tendencies to be prone to ambiguity and misinterpretation. It is often being misunderstood by people from various backgrounds and different levels of knowledge.

A requirement is said to be ambiguous when a same statement is being interpreted differently by different sets of people. A specification is affected by textual ambiguity when it provokes more than one way of reading a statement. Example, "*the customer enters a card and a numeric *personal code. If **it** is not valid then the ATM rejects the card"*. It is ambiguous because the word "*it*" could refer to two distinct objects. It could refer to either *a card* or *a numeric personal code* [1]. Words can be ambiguous in many ways. Linguistic ambiguity can be categorized into several main groups such as semantic, syntactic, pragmatic and lexical [2]. This has been agreed upon and then being enhanced into other types of ambiguity such as coordination ambiguity [3] and anaphoric ambiguity [4], [5].

One of the main reasons for ambiguity is the use of vague words. Words that are being used are not clear and usually lead to more than one meaning. Vagueness can be termed as not clearly expressed, imprecise, ill-defined and lacked expressions [6] . Vagueness shows a boundary of a word's meaning that is not clearly stated [7]. The usage of vague words reduces the level of clarity in a sentence. Vagueness can also be defined as ignorance and absence of knowledge [8]. A vague word can also be defined as a word that has multiple equally good possible candidates of the meaning. When a sentence reaches the 'borderline case' of truth which is neither true nor false, it is considered vague [9]. Malay words such as 'maksimum', 'automatik', 'segera', 'secepat mungkin', 'pantas', 'efisien', 'produktif', 'anggaran', 'kerap' are some of the adjectives considered vague. These words lead to uncertainty and multiple of interpretations and therefore, should be avoided.

Figure 1 below depicts a conceptual view of Malay ambiguity and its' related elements gathered from open-interviews with Malay linguist experts.

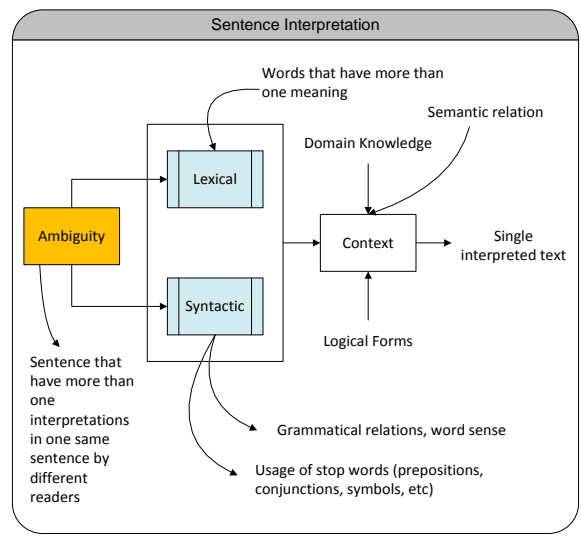

Figure 1: Conceptual view of Malay Ambiguity

## 2. RELATED WORK

Although various researches have focused on disambiguation techniques, not many highlighted how these ambiguous words originated. In addition, most previous researches focused on the English language. Due to limitation in scope, it is quite difficult to refer and construct Malay ambiguous words. Hence, this research is adopting the methods used in English and other languages' methods to suit our area of research.

## 2.1. Vagueness Vs Ambiguity Issues

A sentence must have a unique meaning in order to reflect one's perspective accurately. A sentence containing a vague word, would fail to impart its intended meaning. Vagueness is one of the many sources of ambiguity. For example, "Five piled stones are a heap" [10]. One can consider five piles of stones are a heap, while another might disagree with the statement as he/she may say ten piled stones are then a heap. Vagueness can impact ambiguity that lead to uncertainty and multiple interpretations (refer Figure 2). Vagueness and uncertainty are being distinguished, however, it correlates with one another [9]. They are complimentary but not parallel. Vagueness has a close similarity as semantic indeterminacy or it is termed as 'semantic nihilism' [10]. Therefore, many research concluded that to resolve vagueness, context involvement is necessary [8]. Context is crucial to ensure interpretation is unique in a sentence. [9] Vague can be assigned with different semantic value based on different possible situations, and each of the semantic values is called presification. Vague words leads to imprecise meaning, therefore it triggers ambiguity in a sentence. To disambiguate, we have to go back to its' roots of causal, by eliminating the vague words itself before any ambiguity can be detected.

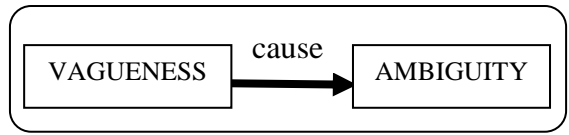

Figure 2: Relationship between Vagueness and Ambiguity

## 2.2. Criterion of Ambiguous Words

A dictionary of 100 ambiguous Arab words that has been developed, takes into consideration more than 10 word senses as the criteria [11]. These senses were extracted from the Arab dictionary. Chantree et al. extracted ambiguous sentences indicate coordination ambiguity and developed ambiguity threshold to set the ambiguity benchmark [12]. Amongst the factors involved in making sure readers understand what a sentence means are sentence length, ambiguous adjectives, adverbs and passive verbs [13]. A list of high potential English ambiguous words has been constructed in an Ambiguity Technical Report as a guideline to avoid ambiguous sentence [14]. Tjong et al. developed rules for clearer sentences in an attempt to avoid ambiguities[15]. These research proof that to begin an investigation to disambiguate an ambiguous sentence, one has to start by determining and identifying the vague words. These vague words could bring misconception and misinterpretation to the readers. As for the writers, they usually are not aware that they are even writing an ambiguous sentence in the first place.

Through previous literatures as guidelines, we have tabled out a criterion of potentially ambiguous words that acts as guidelines to extract the poor words as in Table 1.

Table 1: Criterion of ambiguous words (Malay)

| Criteria | Example (Malay words) |
| --- | --- |
| Words that have more than one word classes | Papar (adj, kk), amat (kk, kt), alam (kk), abstrak (kk, kn) |
| Words have more than one meaning | Perang, semak, alam, akan, |
| Vague adjectives, adverbs and verbs | Efisien, mudah, pantas, segera, lengkap, etc. |
| Words that fall under proposed seven ambiguity | Implicit - efisien Connectives– beberapa |
| attributes: implicit, word class, weakness, temporal, referential and general specific variable. | Weakness – anggaran<br>Temporal – bulanan<br>Referential – sebelum, begini<br>General specific variable – data itu |

## 3. PROPOSED METHODOLOGY

We believed that to minimize and manage ambiguity, one has to go to the root cause. In this case, tracking and identifying the potential vague and ambiguous words are necessary before the process of ambiguity detection can take place. Hence, this strategy will be the first stage from overall of the research work.

### 3.1. Ambiguity Attributes

Table 2 below shows the structure of our proposed Ambiguity Attributes in an attempt to create a list of high potential ambiguous Malay words. These attributes are compiled based on several relevant quality attributes from previous literatures. It consists of six attributes most suitable with Malay words. The ambiguous Malay words are extracted based on these attributes from working RS and some have been translated from English using Dwibahasa Kamus Oxford Fajar [16]. Some of the word class attribute's words were extracted from Kamus Komprehensif Bahasa Melayu [17] for their part of speech (POS).

Table 2. Structure of Ambiguity Attributes

| Ambiguity Attributes | Description |
|---|---|
| Implicit (IMP) : | |
| i. General [18], [14] | Subject or object in the sentence is generic rather than specific. |
| ii. Subjective [18] | Refers to personal opinion or feeling |
| iii. Boundary [14] | It has no definite boundary of true or false (or between yes and no). |
| iv. Unquantifiable [19] | Non-quantifiable |
| Connectives (CON): | |
| i. Adjective[14] | Word belonging to one of the major form classes in any of numerous languages and typically serving as a modifier of a noun to denote a quality of the thing named, to indicate its quantity or extent, or to specify a thing as distinct from something else |
| ii. Adverb [14] | Word belonging to one of the major form classes in any of the numerous languages, typically serving as a modifier of a verb, an adjective, another adverb, a preposition, a phrase, a clause, or a sentence, expressing some relation of manner or quality, place, |

| | time, degree, number, cause, opposition, affirmation, or denial, and in English also serving to connect and to express comment on clause content |
|---|---|
| iii. Verb [14] | Word that characteristically is the grammatical centre of a predicate and expresses an act, occurrence, or mode of being, that in various languages is inflected for agreement with the subject, for tense, for voice, for mood, or for aspect, and that typically has rather full descriptive meaning and characterizing quality but is sometimes nearly devoid of these especially when used as an auxiliary or linking verb |
| iv. Dangling Else [14] | The requirement has no other exit when one case is not met (Exception case) |
| v. Preposition [12], [20] | Connective words. A function word that typically combines with a noun phrase to form a phrase which usually expresses a modification or predication |
| Temporal [19], [14] | Words that has time/duration type that invites multiple interpretation. Un-boundary timing or duration |
| Referential (REF) [14], [19], [4], [5], [21] | Sentence that contains more than one requirement in a sentence. Sentence contains explicit references to (not numbered sentences, not defined, not described, no glossary) |
| Variable (VAR) [14] | Common word that invites vague interpretation and understanding. Too generic. |
| Weakness (WN) [18] | Sentence that contains weak main verb |

### 3.1. Process of creating Malay Ambiguous Lexicons

Figure **3** below depicts the overall process of creating potential ambiguous Malay words repository. Data from sample documents are filtered based on certain criteria. Potentially ambiguous words that have been successfully extracted will undergo testing and verification process before being saved in a repository called Malay Ambiguous Words. The detailed step by step process is described below.

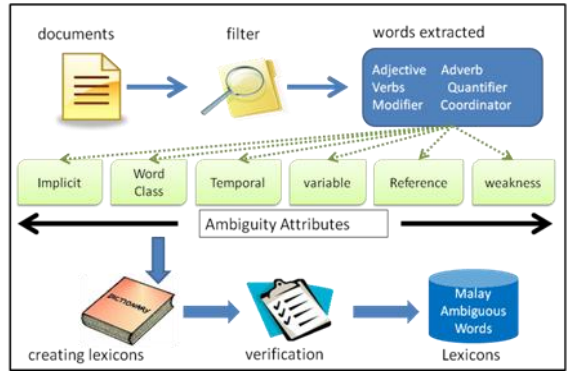

Figure 3. Overall Process of Identifying Ambiguous Words

Step 1: We collected samples of Malay Requirement Specifications from companies as our source of training data. Potentially ambiguous words were extracted from the sentences based on criterion as in Table 1.

From literatures, we constructed six ambiguity attributes that are at most relevant and appropriate with our scope (refer Table 2). Based on these attributes, we mapped the extracted potential Malay ambiguous words with the ambiguity attributes to confirm characteristics of vagueness. By filtering using the above criteria, the list of words considered potentially ambiguous are also referred to as ambiguous candidates. They are kept in a repository to be further analysed using contextual-based detection technique.

Step 3: The identified potentially ambiguous Malay words will undergo a verification process to ensure genuine ambiguity. The verification is expected to be done by Malay linguist experts.

Step 4: The verified words are stored in a database for the next phase of activities.

## 4. DISCUSSION

We have managed to collect 13 sets of Malay language Requirement Specifications from two domains; medical system and student information system. From these sources, a total of 2900 have been words eliminated. Examples of inappropriate words are such as English loanwords, words in short forms, double words such as 'rekod-rekod', 'kata nama khas (KNK)' and symbols such as full stops and other symbols. We then managed to extract 120 potentially ambiguous Malay words. Table 3 below is the statistics of the words' mapping onto their appropriate Ambiguity Attributes.

Table 3. Words mapping based on Ambiguity Attributes

|     | IMP  | CON  | T   | REF  | VAR  | WN   |
| --- | ---- | ---- | --- | ---- | ---- | ---- |
| Tot | 51   | 41   | 11  | 27   | 22   | 21   |
| %   | 42.5 | 34.2 | 9.2 | 22.5 | 18.3 | 17.5 |

From the statistic generated, the highest percentage of potential ambiguous Malay words falls under 'Implicit' category followed by 'Connectives' category and 'Referential'. The articulated data shows that potentially ambiguous Malay words most used are very generic, has a vague boundary, too subjective and reflects an unquantifiable criterion. These are the normal reason that triggers ambiguity. The list of ambiguous words is currently undergoing a verification process by Malay linguist experts. Two experts with the relevant background and expertise of the domain were selected from Faculty of Communication and Malay Language (FKBM), Universiti Putra Malaysia (UPM).

## 5. CONCLUSION

In system requirement, linguistic ambiguity is often ignored or mistakenly unacknowledged. This leads to misunderstanding from both users and system developer's side, thereby contributing towards a failed system. The after effect of the situation could jeopardize system development cycle and project's time limitation as well as budgets. The Malay requirement specification environment still lacks in research that focussed on this situation. We have presented here a method to identify potential ambiguous Malay words and managed to construct a list of 120 potential commonly used ambiguous Malay words in a Malay requirement specification. This study is an attempt to assist writers to avoid using the high potential ambiguous words and promote greater clarity in sentence construction of documentation and significantly reduce misinterpretation by readers.

## ACKNOWLEDGEMENTS

We would like to thank the Malay Linguist Experts from UPM, UM and UKM for their support and helpful comments. This work is supported and funded by PhD scholarship from Universiti Utara Malaysia and Universiti Putra Malaysia.

**Authors**

Hazlina Haron is a PhD student in the Faculty of Computer Science and Information Technology, Universiti Putra Malaysia. She holds a B.Sc in Information Technology (1998) from Universiti Utara Malaysia and M.Sc in Computer Science (2006) from Universiti Putra Malaysia. Currently she is a tutor in Universiti Utara Malaysia. She had 8 years of experience as Software Engineer while working at Telekom Applied Business Sdn Bhd, a subsidiary of Telekom Malaysia Berhad and 3 years of experience as Assistant Manager doing Project Management Office at Telekom Malaysia Berhad. Her studies involve developing a technique best suited to detect ambiguity in Malay natural language. Her research areas include requirement engineering, natural language processing and ambiguity.

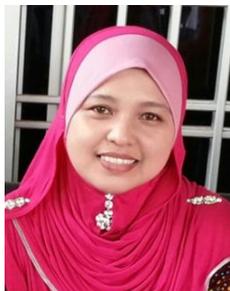

Abdul Azim Abd Ghani is a Professor in the Department of Software Engineering and Information Technology, Universiti Putra Malaysia. He received B.Sc in Mathematics/Computer Science from Indiana State University in 1984 and M.Sc in Computer Science from University of Miami in 1985. He received the Ph.D. in Software Engineering from University of Strathclyde in 1993. His research interests are software engineering, software measurement, software quality, and security in computing.

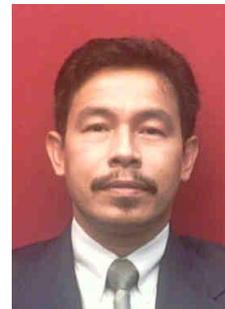